# A new integrated symmetrical table for genetic codes

**JIAN-JUN SHU**
School of Mechanical & Aerospace Engineering,
Nanyang Technological University, 50 Nanyang Avenue, Singapore 639798

## Abstract

Degeneracy is a salient feature of genetic codes, because there are more codons than amino acids. The conventional table for genetic codes suffers from an inability of illustrating a symmetrical nature among genetic base codes. In fact, because the conventional wisdom avoids the question, there is little agreement as to whether the symmetrical nature actually even exists. A better understanding of symmetry and an appreciation for its essential role in the genetic code formation can improve our understanding of nature's coding processes. Thus, it is worth formulating a new integrated symmetrical table for genetic codes, which is presented in this paper. It could be very useful to understand the Nobel laureate Crick's wobble hypothesis — how one transfer ribonucleic acid can recognize two or more synonymous codons, which is an unsolved fundamental question in biological science.

## 1 Introduction

The discovery of double-helix molecular structure of deoxyribonucleic acid (DNA) by Watson and Crick (1953) is one of landmarks in the history of science. It represents the birth of molecular biology. On the cellular level, the living organisms are classified into prokaryotes and eukaryotes. The prokaryotes are unicellular life forms while the eukaryotes include human, animal and fungus. All prokaryotic and eukaryotic cells share a common process by which information encoded by a gene is used to produce the corresponding protein. This process is called protein biosynthesis and accomplished in two steps: transcription and translation.

During transcription, DNA is transcribed into ribonucleic acid (RNA). DNA carries the genetic information, while RNA is used to synthesize proteins. DNA consists of a strand of bases, namely Adenine (A), Thymine (T), Guanine (G) and Cytosine (C), whereas RNA has A, G, C and Uracil (U) instead of T. Then, translation occurs where proteins (molecules composed of a long chain of amino acids) are built upon the codons in RNA. Each codon, which is a set of three adjoined nucleotides (triplet), specifies one amino acid or termination signal (Crick *et al*., 1961).

There are 20 amino acids, namely Histidine (His/H), Arginine (Arg/R), Lysine (Lys/K), Phenylalanine (Phe/F), Alanine (Ala/A), Leucine (Leu/L), Methionine (Met/M), Isoleucine (Ile/I), Tryptophan (Trp/W), Proline (Pro/P), Valine (Val/V),

Cysteine (Cys/C), Glycine (Gly/G), Glutamine (Gln/Q), Asparagine (Asn/N), Serine (Ser/S), Tyrosine (Tyr/Y), Threonine (Thr/T), Aspartic acid (Asp/D) and Glutamic acid (Glu/E). For the formation of proteins in living organism cells, it is found that each amino acid can be specified by either a minimum of one codon or up to a maximum of six possible codons. In other words, different codons specify the different number of amino acids. A table for genetic codes is a representation of translation for illustrating the different amino acids with their respectively specifying codons, that is, a set of rules by which information encoded in genetic material (RNA sequences) is translated into proteins (amino acid sequences) by living cells. There are a total of 64 possible codons, but there are only 20 amino acids specified by them. Therefore, degeneracy is a salient feature of genetic codes. Genetic information is stored in DNA in the form of sequences of nucleotides which is made clearly in the double-helix model, but it does not provide any clue on how one transfer ribonucleic acid (tRNA) can recognize two or more synonymous codons. Therefore, deciphering the genetic codes becomes a problem. Up to now, it is still unable to find out the reason or explanation for these kinds of characteristics and relationships between codons and amino acids. Therefore, it has always been an interesting area for us to explore and obtain any explanation further.

The table for genetic codes allows us to identify a codon and the individual amino acid assigned to the codon by nature. These assignment tables may come in a variety of forms, but they all suffer from an inability of illustrating a symmetrical nature among genetic base codes. In fact, because the conventional wisdom avoids the question, there is little agreement as to whether the symmetrical nature actually even exists. A better understanding of symmetry and an appreciation for its essential role in the genetic code formation can improve our understanding of nature's coding processes. Thus, it is worth formulating a new integrated symmetrical table for genetic codes.

## 2 Genetic codes

The genetic codes for translation can be categorized into two main categories: nuclear and mitochondrial codes, which are the genetic codes of nuclear deoxyribonucleic acid (nDNA) and mitochondrial deoxyribonucleic acid (mtDNA) respectively. Each category has various different genetic codes for the translation of a particular class, genus or species of living organisms. Not all organisms can use standard nuclear code for translation and some organisms of the same family can have the different set of translation codes. As shown in Figure 1, there are total 16 genetic codes, that is, standard nuclear, bacterial, archaeal & plant plastid code (NM1) (Nirenberg and Matthaei, 1961), mold, protozoan, coelenterate mitochondrial & *mycoplasma*/*spiroplasma* nuclear code (NM2) (Fox, 1987), euplotid nuclear code (N1) (Hoffman *et al*., 1995), *blepharisma* nuclear code (N2) (Liang and Heckmann, 1993), ciliate, dasycladacean & *hexamita* nuclear code (N3) (Schneider *et al*., 1989), alternative yeast nuclear code (N4) (Ohama *et al*., 1993), vertebrate mitochondrial code (M1) (Barrell *et al*., 1979), invertebrate mitochondrial code (M2) (Batuecas *et al*., 1988), ascidian mitochondrial code (M3) (Yokobori *et al*., 1993), echinoderm & flatworm mitochondrial code (M4) (Himeno *et al*., 1987), alternative flatworm mitochondrial code (M5) (Bessho *et al*., 1992), trematode mitochondrial code (M6) (Garey and Wolstenholme, 1989), chlorophycean mitochondrial code (M7) (Hayashi-Ishimaru *et al*., 1996), *thraustochytrium* mitochondrial code (M8) (Goldstein, 1973), *scenedesmus obliquus* mitochondrial code (M9) (Nedelcu *et al*., 2000) and yeast mitochondrial code (M10) (Clark-Walker and Weiller, 1994).

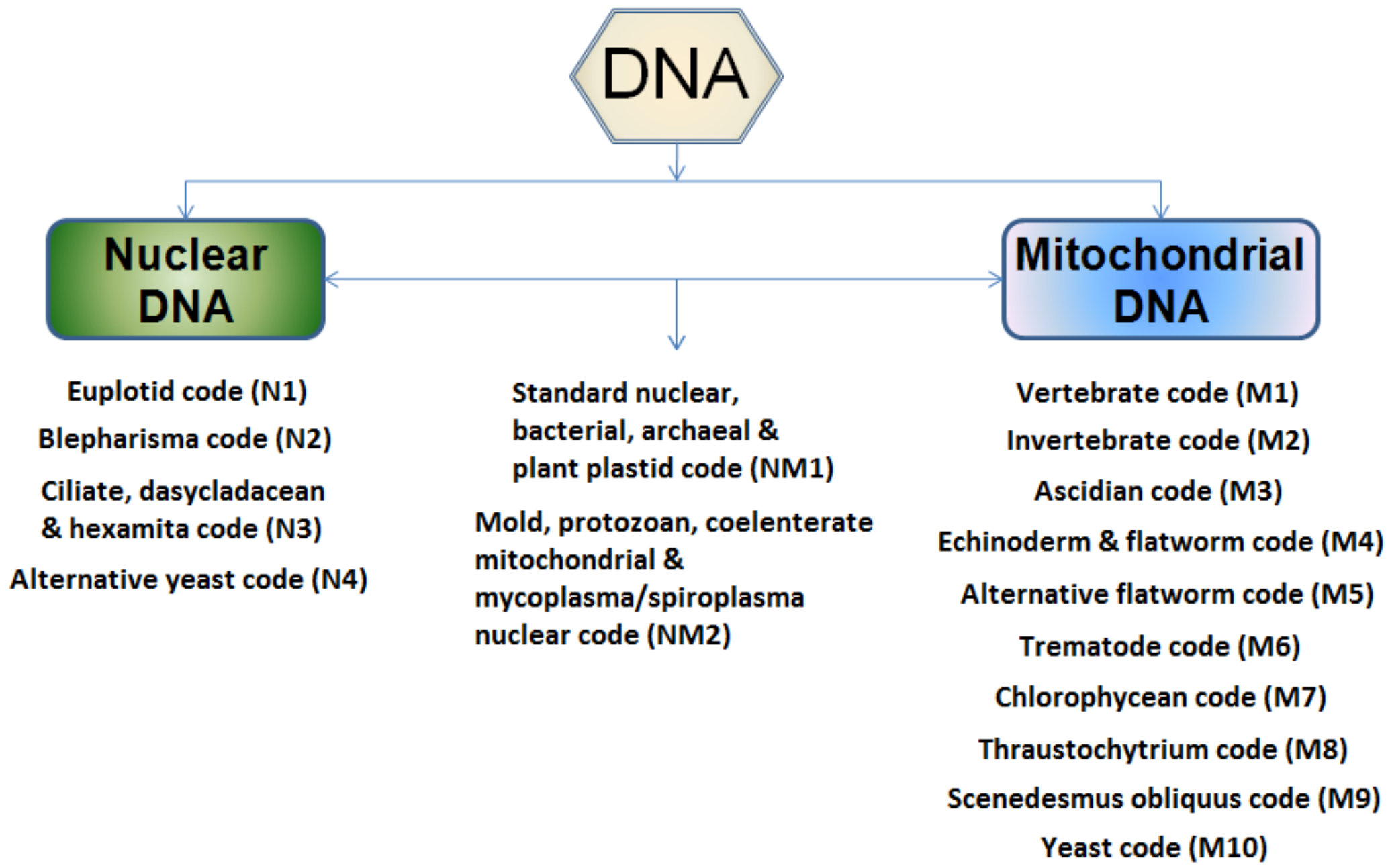

Figure 1: Genetic codes

## 3 Symmetrical genetic codes

In standard nuclear code (Nirenberg and Matthaei, 1961), the arrangement of amino acid assignment is not random, presumably as the product of evolution to enhance stability in the face of mutation (Freeland and Hurst, 1998; Freeland *et al*., 2000, 2003; Sella and Ardell, 2006), tRNA misloading (Yang, 2004; Jestin and Soulé, 2007; Seligmann, 2010b, 2011, 2012), frame shift (Seligmann and Pollock, 2004; Itzkovitz and Alon, 2007; Seligmann, 2007, 2010a) and protein misfolding (Guilloux and Jestin, 2012). As shown in Figure 2(a), the original genetic code is arranged in the conventional form following the mapping sequence from left to right. The appearance of degeneracy in the conventional table implies the existence of certain symmetry for codon multiplicity assignment (Findley *et al*., 1982; Shcherbak, 1988; Bashford *et al*., 1998; Hornos *et al*., 2004; Nikolajewa *et al*., 2006; Gavish *et al*., 2007; Rosandić and Paar, 2014). Figure 2(b) shows another way of arrangement of standard nuclear code by changing sequence position from **“1-2-3”** to **“1-3-2”** and base sequence at each position from **“U-C-A-G”** to **“C-A-G-U”**. This is different from the conventional table of standard nuclear code. A newly-formulated genetic code presents a new perspective of genetic code.

**NM1**

| 1st Position | | 2nd Position: U | C | A | G | 3rd Position |
|---|---|---|---|---|---|---|
| | U | F | S | Y | C | U |
| | U | F | S | Y | C | C |
| | U | L | S | * | * | A |
| | U | L | S | * | W | G |
| | C | L | P | H | R | U |
| | C | L | P | H | R | C |
| | C | L | P | Q | R | A |
| | C | L | P | Q | R | G |
| | A | I | T | N | S | U |
| | A | I | T | N | S | C |
| | A | I | T | K | R | A |
| | A | M | T | K | R | G |
| | G | V | A | D | G | U |
| | G | V | A | D | G | C |
| | G | V | A | E | G | A |
| | G | V | A | E | G | G |

(a) original table

**NM1**

| 1st Position | | 3rd Position: C | A | G | U | 2nd Position |
|---|---|---|---|---|---|---|
| | C | P | P | P | P | C |
| | C | H | Q | Q | H | A |
| | C | R | R | R | R | G |
| | C | L | L | L | L | U |
| | A | T | T | T | T | C |
| | A | N | K | K | N | A |
| | A | S | R | R | S | G |
| | A | I | I | M | I | U |
| | U | S | S | S | S | C |
| | U | Y | * | * | Y | A |
| | U | C | * | W | C | G |
| | U | F | L | L | F | U |
| | G | A | A | A | A | C |
| | G | D | E | E | D | A |
| | G | G | G | G | G | G |
| | G | V | V | V | V | U |

(b) symmetrical table

Figure 2: Standard nuclear code

From Figure 2(b), it shows that the rearranged standard nuclear code exhibits a symmetrical pattern along the vertical centerline except that the 4 codons (UGA||UGG and AUA||AUG) at the center of table are not symmetrical to each other.

Although the 4 codons are not perfectly mirrored about the vertical centerline, it is still possible to see the pairing characteristics between "C|U"-pyrimidine and "A|G"-purine at the third position. The third (or wobble) base plays a significant role in encoding the type of amino acid.

The rearrangement of other genetic codes for both nDNA and mtDNA also exhibits such a symmetrical pattern. As shown in Figure 3, the majority of codon pairs are symmetric about the vertical centerline. However there are still some codon pairs that are asymmetric. The four rearranged genetic codes, namely, vertebrate (M1), invertebrate (M2), ascidian (M3) and yeast (M10), exhibit a perfect symmetrical pattern along the vertical centerline and all belong to the mitochondrial genetic family. It is worth pointing out, that the only perfect symmetry of vertebrate mitochondrial code (M1) was noted (Lehmann, 2000; Gonzalez *et al*., 2013).

**NM1**

| | | 3rd Position | | | |
|---|---|---|---|---|---|
| 1st Position | 2nd Position | C | A | G | U |
| C | C | P | P | P | P |
| C | A | H | Q | Q | H |
| C | G | R | R | R | R |
| C | U | L | L | L | L |
| A | C | T | T | T | T |
| A | A | N | K | K | N |
| A | G | S | R | R | S |
| A | U | I | I | M | I |
| U | C | S | S | S | S |
| U | A | Y | * | * | Y |
| U | G | C | * | W | C |
| U | U | F | L | L | F |
| G | C | A | A | A | A |
| G | A | D | E | E | D |
| G | G | G | G | G | G |
| G | U | V | V | V | V |

**NM2**

| | | 3rd Position | | | |
|---|---|---|---|---|---|
| 1st Position | 2nd Position | C | A | G | U |
| C | C | P | P | P | P |
| C | A | H | Q | Q | H |
| C | G | R | R | R | R |
| C | U | L | L | L | L |
| A | C | T | T | T | T |
| A | A | N | K | K | N |
| A | G | S | R | R | S |
| A | U | I | I | M | I |
| U | C | S | S | S | S |
| U | A | Y | * | * | Y |
| U | G | C | W | W | C |
| U | U | F | L | L | F |
| G | C | A | A | A | A |
| G | A | D | E | E | D |
| G | G | G | G | G | G |
| G | U | V | V | V | V |

**N1**

| | | 3rd Position | | | |
|---|---|---|---|---|---|
| 1st Position | 2nd Position | C | A | G | U |
| C | C | P | P | P | P |
| C | A | H | Q | Q | H |
| C | G | R | R | R | R |
| C | U | L | L | L | L |
| A | C | T | T | T | T |
| A | A | N | K | K | N |
| A | G | S | R | R | S |
| A | U | I | I | M | I |
| U | C | S | S | S | S |
| U | A | Y | * | * | Y |
| U | G | C | C | W | C |
| U | U | F | L | L | F |
| G | C | A | A | A | A |
| G | A | D | E | E | D |
| G | G | G | G | G | G |
| G | U | V | V | V | V |

**N2**

| | | 3rd Position | | | |
|---|---|---|---|---|---|
| 1st Position | 2nd Position | C | A | G | U |
| C | C | P | P | P | P |
| C | A | H | Q | Q | H |
| C | G | R | R | R | R |
| C | U | L | L | L | L |
| A | C | T | T | T | T |
| A | A | N | K | K | N |
| A | G | S | R | R | S |
| A | U | I | I | M | I |
| U | C | S | S | S | S |
| U | A | Y | * | Q | Y |
| U | G | C | * | W | C |
| U | U | F | L | L | F |
| G | C | A | A | A | A |
| G | A | D | E | E | D |
| G | G | G | G | G | G |
| G | U | V | V | V | V |

**N3**

| | | 3rd Position | | | |
|---|---|---|---|---|---|
| 1st Position | 2nd Position | C | A | G | U |
| C | C | P | P | P | P |
| C | A | H | Q | Q | H |
| C | G | R | R | R | R |
| C | U | L | L | L | L |
| A | C | T | T | T | T |
| A | A | N | K | K | N |
| A | G | S | R | R | S |
| A | U | I | I | M | I |
| U | C | S | S | S | S |
| U | A | Y | Q | Q | Y |
| U | G | C | * | W | C |
| U | U | F | L | L | F |
| G | C | A | A | A | A |
| G | A | D | E | E | D |
| G | G | G | G | G | G |
| G | U | V | V | V | V |

**N4**

| | | 3rd Position | | | |
|---|---|---|---|---|---|
| 1st Position | 2nd Position | C | A | G | U |
| C | C | P | P | P | P |
| C | A | H | Q | Q | H |
| C | G | R | R | R | R |
| C | U | L | L | S | L |
| A | C | T | T | T | T |
| A | A | N | K | K | N |
| A | G | S | R | R | S |
| A | U | I | I | M | I |
| U | C | S | S | S | S |
| U | A | Y | * | * | Y |
| U | G | C | * | W | C |
| U | U | F | L | L | F |
| G | C | A | A | A | A |
| G | A | D | E | E | D |
| G | G | G | G | G | G |
| G | U | V | V | V | V |

**M1**

| | | 3rd Position | | | |
|---|---|---|---|---|---|
| 1st Position | 2nd Position | C | A | G | U |
| C | C | P | P | P | P |
| C | A | H | Q | Q | H |
| C | G | R | R | R | R |
| C | U | L | L | L | L |
| A | C | T | T | T | T |
| A | A | N | K | K | N |
| A | G | S | * | * | S |
| A | U | I | M | M | I |
| U | C | S | S | S | S |
| U | A | Y | * | * | Y |
| U | G | C | W | W | C |
| U | U | F | L | L | F |
| G | C | A | A | A | A |
| G | A | D | E | E | D |
| G | G | G | G | G | G |
| G | U | V | V | V | V |

**M2**

| | | 3rd Position | | | |
|---|---|---|---|---|---|
| 1st Position | 2nd Position | C | A | G | U |
| C | C | P | P | P | P |
| C | A | H | Q | Q | H |
| C | G | R | R | R | R |
| C | U | L | L | L | L |
| A | C | T | T | T | T |
| A | A | N | K | K | N |
| A | G | S | S | S | S |
| A | U | I | M | M | I |
| U | C | S | S | S | S |
| U | A | Y | * | * | Y |
| U | G | C | W | W | C |
| U | U | F | L | L | F |
| G | C | A | A | A | A |
| G | A | D | E | E | D |
| G | G | G | G | G | G |
| G | U | V | V | V | V |

**M3**

| 1st Position | 2nd Position | 3rd Position C | 3rd Position A | 3rd Position G | 3rd Position U |
|---|---|---|---|---|---|
| C | C | P | P | P | P |
| C | A | H | Q | Q | H |
| C | G | R | R | R | R |
| C | U | L | L | L | L |
| A | C | T | T | T | T |
| A | A | N | K | K | N |
| A | G | S | G | G | S |
| A | U | I | M | M | I |
| U | C | S | S | S | S |
| U | A | Y | * | * | Y |
| U | G | C | W | W | C |
| U | U | F | L | L | F |
| G | C | A | A | A | A |
| G | A | D | E | E | D |
| G | G | G | G | G | G |
| G | U | V | V | V | V |

**M4**

| 1st Position | 2nd Position | 3rd Position C | 3rd Position A | 3rd Position G | 3rd Position U |
|---|---|---|---|---|---|
| C | C | P | P | P | P |
| C | A | H | Q | Q | H |
| C | G | R | R | R | R |
| C | U | L | L | L | L |
| A | C | T | T | T | T |
| A | A | N | N | K | N |
| A | G | S | S | S | S |
| A | U | I | I | M | I |
| U | C | S | S | S | S |
| U | A | Y | * | * | Y |
| U | G | C | W | W | C |
| U | U | F | L | L | F |
| G | C | A | A | A | A |
| G | A | D | E | E | D |
| G | G | G | G | G | G |
| G | U | V | V | V | V |

**M5**

| 1st Position | 2nd Position | 3rd Position C | 3rd Position A | 3rd Position G | 3rd Position U |
|---|---|---|---|---|---|
| C | C | P | P | P | P |
| C | A | H | Q | Q | H |
| C | G | R | R | R | R |
| C | U | L | L | L | L |
| A | C | T | T | T | T |
| A | A | N | N | K | N |
| A | G | S | S | S | S |
| A | U | I | I | M | I |
| U | C | S | S | S | S |
| U | A | Y | Y | * | Y |
| U | G | C | W | W | C |
| U | U | F | L | L | F |
| G | C | A | A | A | A |
| G | A | D | E | E | D |
| G | G | G | G | G | G |
| G | U | V | V | V | V |

**M6**

| 1st Position | 2nd Position | 3rd Position C | 3rd Position A | 3rd Position G | 3rd Position U |
|---|---|---|---|---|---|
| C | C | P | P | P | P |
| C | A | H | Q | Q | H |
| C | G | R | R | R | R |
| C | U | L | L | L | L |
| A | C | T | T | T | T |
| A | A | N | N | K | N |
| A | G | S | S | S | S |
| A | U | I | M | M | I |
| U | C | S | S | S | S |
| U | A | Y | * | * | Y |
| U | G | C | W | W | C |
| U | U | F | L | L | F |
| G | C | A | A | A | A |
| G | A | D | E | E | D |
| G | G | G | G | G | G |
| G | U | V | V | V | V |

**M7**

| 1st Position | 2nd Position | 3rd Position C | 3rd Position A | 3rd Position G | 3rd Position U |
|---|---|---|---|---|---|
| C | C | P | P | P | P |
| C | A | H | Q | Q | H |
| C | G | R | R | R | R |
| C | U | L | L | L | L |
| A | C | T | T | T | T |
| A | A | N | K | K | N |
| A | G | S | R | R | S |
| A | U | I | I | M | I |
| U | C | S | S | S | S |
| U | A | Y | * | L | Y |
| U | G | C | * | W | C |
| U | U | F | L | L | F |
| G | C | A | A | A | A |
| G | A | D | E | E | D |
| G | G | G | G | G | G |
| G | U | V | V | V | V |

**M8**

| 1st Position | 2nd Position | 3rd Position C | 3rd Position A | 3rd Position G | 3rd Position U |
|---|---|---|---|---|---|
| C | C | P | P | P | P |
| C | A | H | Q | Q | H |
| C | G | R | R | R | R |
| C | U | L | L | L | L |
| A | C | T | T | T | T |
| A | A | N | K | K | N |
| A | G | S | R | R | S |
| A | U | I | I | M | I |
| U | C | S | S | S | S |
| U | A | Y | * | * | Y |
| U | G | C | * | W | C |
| U | U | F | * | L | F |
| G | C | A | A | A | A |
| G | A | D | E | E | D |
| G | G | G | G | G | G |
| G | U | V | V | V | V |

**M9**

| 1st Position | 2nd Position | 3rd Position C | 3rd Position A | 3rd Position G | 3rd Position U |
|---|---|---|---|---|---|
| C | C | P | P | P | P |
| C | A | H | Q | Q | H |
| C | G | R | R | R | R |
| C | U | L | L | L | L |
| A | C | T | T | T | T |
| A | A | N | K | K | N |
| A | G | S | R | R | S |
| A | U | I | I | M | I |
| U | C | S | * | S | S |
| U | A | Y | * | L | Y |
| U | G | C | * | W | C |
| U | U | F | L | L | F |
| G | C | A | A | A | A |
| G | A | D | E | E | D |
| G | G | G | G | G | G |
| G | U | V | V | V | V |

**M10**

| 1st Position | 2nd Position | 3rd Position C | 3rd Position A | 3rd Position G | 3rd Position U |
|---|---|---|---|---|---|
| C | C | P | P | P | P |
| C | A | H | Q | Q | H |
| C | G | R | R | R | R |
| C | U | T | T | T | T |
| A | C | T | T | T | T |
| A | A | N | K | K | N |
| A | G | S | R | R | S |
| A | U | I | M | M | I |
| U | C | S | S | S | S |
| U | A | Y | * | * | Y |
| U | G | C | W | W | C |
| U | U | F | L | L | F |
| G | C | A | A | A | A |
| G | A | D | E | E | D |
| G | G | G | G | G | G |
| G | U | V | V | V | V |

Figure 3: Rearranged codes

# 4 An integrated table for genetic codes

In view of the symmetrical and asymmetrical characteristics of all 16 rearranged genetic codes, there is a question that many may raise whether a perfect symmetrical genetic code is the origin or ultimate product of evolutionary progress. The total 16 sets of genetic codes that are used in different biological species may give us the clue of how the evolutionary process happened. Today various species are quite different in terms of appearance, but they may have the same ancestor. Therefore, a new integrated table for genetic codes is needed to discover any possible regularity occurring in all genetic codes. As shown in Figure 4, the integrated table for genetic codes has a standardized template in the center which is identical for all genetic codes and surrounded with the unique signature of each rearranged genetic code. The proposed integrated table for genetic codes aims to provide the user with a comprehensive and simple genetic translation interface, which is comprised of the entire different genetic translation codes. Simply by replacing the blank center of standardized template with the surrounding unique signature, the user can then obtain desired genetic translation table for each organism.

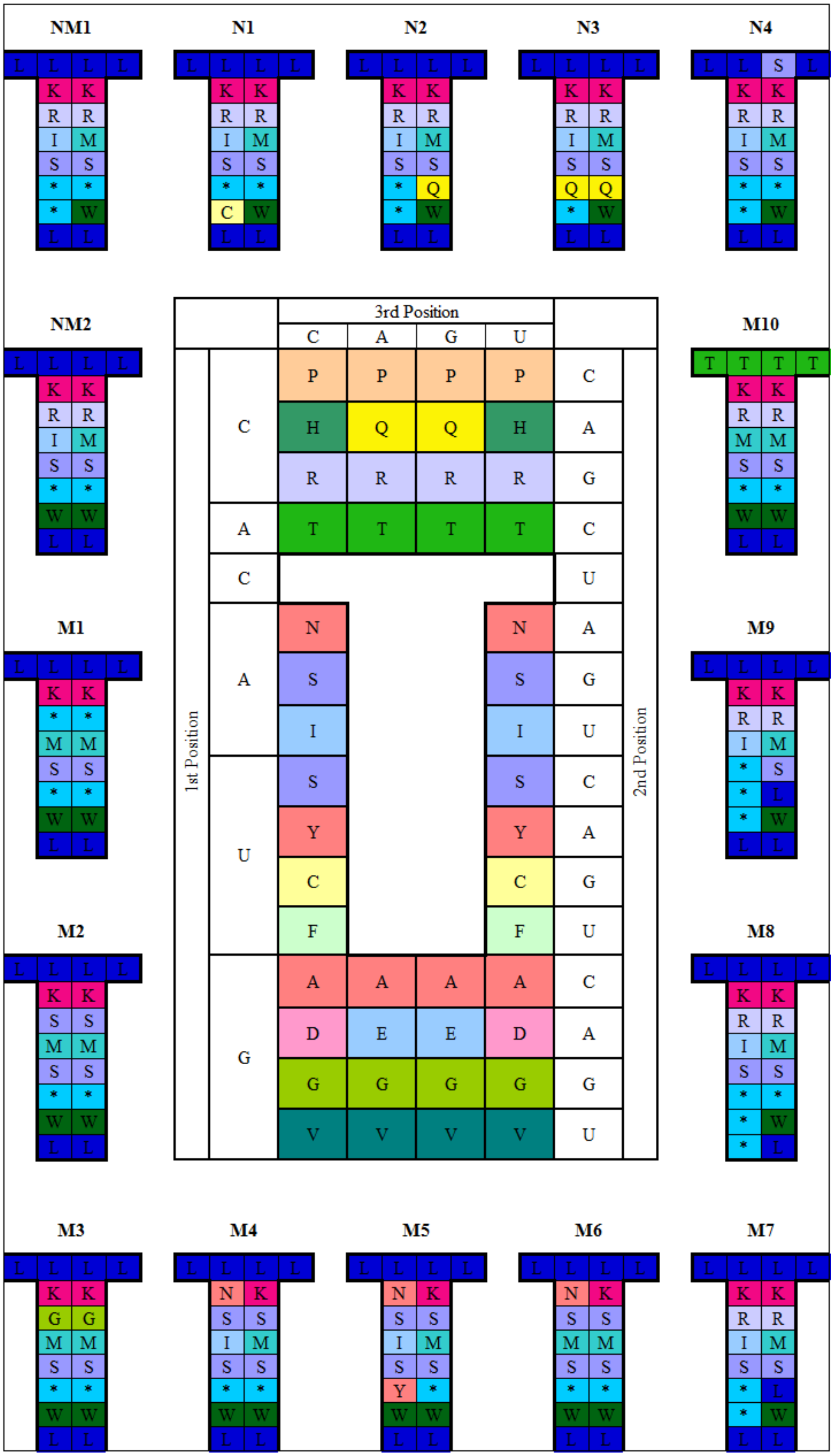

Figure 4: An integrated table for genetic codes

# 5 Concluding remarks

As shown in Figure 4, the STOP (*) codon of standard nuclear code is mutated to other amino acids in almost every non-standard code. The only two that do not contain STOP (*) codon mutation are alternative yeast nuclear code (N4) and *thraustochytrium* mitochondrial code (M8). This seems to imply that, the STOP (*) codon is the most unstable codon in genetic codes, or could be seen as an empty shell, which could easily be replaced by the nearby amino acids. It is worth mentioning to this end that the five genetic codes, euplotid nuclear code (N1), alternative yeast nuclear code (N4), echinoderm & flatworm mitochondrial code (M4), alternative flatworm mitochondrial code (M5) and trematode mitochondrial code (M6), do not follow the intuition based on standard nuclear code (Nirenberg and Matthaei, 1961), which asymmetry is restricted to the 'punctuation' codons, START (Met/M) and STOP (*) codons (Rumer, 1966; Shcherbak, 1989; Kozyrev and Khrennikov, 2010; Rosandić *et al*., 2013; Seligmann, 2015).

There is the existence of symmetrical and asymmetrical characteristics in genetic codes. The presence of symmetry enables the ease/efficiency of live formation and the role of symmetry-breaking (asymmetry) is to enable evolution/adaptation to take place (Elitzur, 1997; Seligmann, 2000; Antoneli and Forger, 2011; Lenstra, 2014). In short, lives are formed due to symmetry but evolved due to asymmetry.

The newly-formulated integrated symmetrical table provides the new perspective of possible codon-amino acid relationship and explains the hidden meaning/logic behind the degenerative genetic codes, and can be used to build programmable biomolecular mediated processors (Shu *et al*., 2011, 2015, 2016; Wong *et al*., 2015) for efficient genome editing (Ishino *et al*., 1987; Jinek *et al*., 2012) by taking both symmetrical and asymmetrical characteristics into account.

# References

Antoneli, F., Forger, M., 2011. Symmetry breaking in the genetic code: Finite groups. *Mathematical and Computer Modelling* **53**(7-8), 1469–1488.

Barrell, B.G., Bankier, A.T., Drouin, J., 1979. A different genetic code in human mitochondria. *Nature* **282**(5735), 189–194.

Bashford, J.D., Tsohantjis, I., Jarvis, P.D., 1998. A supersymmetric model for the evolution of the genetic code. *Proceedings of the National Academy of Sciences of the United States of America* **95**(3), 987–992.

Batuecas, B., Garesse, R., Calleja, M., Valverde, J.R., Marco, R., 1988. Genome organization of *Artemia* mitochondrial DNA. *Nucleic Acids Research* **16**(14A), 6515–6529.

Bessho, Y., Ohama, T., Osawa, S., 1992. Planarian mitochondria II. The unique genetic code as deduced from cytochrome c oxidase subunit I gene sequences. *Journal of Molecular Evolution* **34**(4), 331–335.

Clark-Walker, G.D., Weiller, G.F., 1994. The structure of the small mitochondrial DNA of Kluyveromyces thermotolerans is likely to reflect the ancestral gene order in fungi. *Journal of Molecular Evolution* **38**(6), 593–601.

Crick, F.H.C., Barnett, L., Brenner, S., Watts-Tobin, R.J., 1961. General nature of the genetic code for proteins. *Nature* **192**(4809), 1227–1232.

Elitzur, A.C., 1997. Constancy, uniformity and symmetry of living systems: The computational functions of morphological invariance. *Biosystems* **43**(1), 41–53.

Findley, G.L., Findley, A.M., McGlynn, S.P., 1982. Symmetry characteristics of the genetic code. *Proceedings of the National Academy of Sciences of the United States of America-Physical Sciences* **79**(22), 7061–7065.

Fox, T.D., 1987. Natural variation in the genetic code. *Annual Review of Genetics* **21**, 67–91.

Freeland, S.J., Hurst, L.D., 1998. The genetic code is one in a million. *Journal of Molecular Evolution* **47**(3), 238–248.

Freeland, S.J., Knight, R.D., Landweber, L.F., Hurst, L.D., 2000. Early fixation of an optimal genetic code. *Molecular Biology and Evolution* **17**(4), 511–518.

Freeland, S.J., Wu, T., Keulmann, N., 2003. The case for an error minimizing standard genetic code. *Origins of Life and Evolution of the Biospheres* **33**(4), 457–477.

Garey, J.R., Wolstenholme, D.R., 1989. Platyhelminth mitochondrial DNA: Evidence for early evolutionary origin of a $tRNA^{ser}AGN$ that contains a dihydrouridine arm replacement loop, and of serine-specifying AGA and AGG codons. *Journal of Molecular Evolution* **28**(5), 374–387.

Gavish, M. Peled, A., Chor, B., 2007. Genetic code symmetry and efficient design of GC-constrained coding sequences. *Bioinformatics* **23**(2), E57–E63.

Goldstein, S., 1973. Zoosporic marine fungi (Thraustochytriaceae and Dermocystidiaceae). *Annual Review of Microbiology* **27**, 13–26.

Gonzalez, D.L., Giannerini, S., Rosa, R., 2013. On the origin of the mitochondrial genetic code: Towards a unified mathematical framework for the management of genetic information. *Nature Precedings* 1–20.

Guilloux, A., Jestin, J.-L., 2012. The genetic code and its optimization for kinetic energy conservation in polypeptide chains. *Biosystems* **109**(2), 141–144.

Hayashi-Ishimaru, Y., Ohama, T., Kawatsu, Y., Nakamura, K., Osawa, S., 1996. UAG is a sense codon in several chlorophycean mitochondria. *Current Genetics* **30**(1), 29–33.

Himeno, H., Masaki, H., Kawai, T., Ohta, T., Kumagai, I., Miura, K., Watanabe, K., 1987. Unusual genetic codes and a novel gene structure for $tRNA^{Ser}AGY$ in starfish mitochondrial DNA. *Gene* **56**(2-3), 219–230.

Hoffman, D.C., Anderson, R.C., DuBois, M.L., Prescott, D.M., 1995. Macronuclear gene-sized molecules of hypotrichs. *Nucleic Acids Research* **23**(8), 1279–1283.

Hornos, J.E.M., Braggion, L., Magini, M., Forger, M., 2004. Symmetry preservation in the evolution of the genetic code. *IUBMB Life* **56**(3), 125–130.

Ishino, Y., Shinagawa, H., Makino, K., Amemura, M., Nakata, A., 1987. Nucleotide sequence of the *iap* gene, responsible for alkaline phosphatase isozyme conversion in *Escherichia-coli*, and identification of the gene product. *Journal of Bacteriology* **169**(12), 5429–5433.

Itzkovitz, S., Alon, U., 2007. The genetic code is nearly optimal for allowing additional information within protein-coding sequences. *Genome Research* **17**(4), 405–412.

Jestin, J.-L., Soulé, C., 2007. Symmetries by base substitutions in the genetic code predict 2' or 3' aminoacylation of tRNAs. *Journal of Theoretical Biology* **247**(2), 391–394.

Jinek, M., Chylinski, K., Fonfara, I., Hauer, M., Doudna, J.A., Charpentier, E., 2012. A programmable dual-RNA-guided DNA endonuclease in adaptive bacterial immunity. *Science* **337**(6096), 816–821.

Kozyrev, S.V., Khrennikov, A.Y., 2010. 2-Adic numbers in genetics and Rumer's symmetry. *Doklady Mathematics* 81(1), 128–130.

Lehmann, J., 2000. Physico-chemical constraints connected with the coding properties of the genetic system. *Journal of Theoretical Biology* **202**(2), 129–144.

Lenstra, R., 2014. Evolution of the genetic code through progressive symmetry breaking. *Journal of Theoretical Biology* **347**, 95–108.

Liang, A., Heckmann, K., 1993. *Blepharisma* uses UAA as a termination codon. *Naturwissenschaften* **80**(5), 225–226.

Nedelcu, A.M., Lee, R.W., Lemieux, C., Gray, M.W., Burger, G., 2000. The complete mitochondrial DNA sequence of *Scenedesmus obliquus* reflects an intermediate stage in the evolution of the green algal mitochondrial genome. *Genome Research* **10**(6), 819–831.

Nikolajewa, S., Friedel, M., Beyer, A., Wilhelm, T., 2006. The new classification scheme of the genetic code, its early evolution, and tRNA usage. *Journal of Bioinformatics and Computational Biology* **4**(2), 609–620.

Nirenberg, M.W., Matthaei, J.H., 1961. The dependence of cell-free protein synthesis in *E. coli* upon naturally occurring or synthetic polyribonucleotides. *Proceedings of the National Academy of Sciences of the United States of America* **47**(10), 1588–1602.

Ohama, T., Suzuki, T., Mori, M., Osawa, S., Ueda, T., Watanabe, K., Nakase, T., 1993. Non-universal decoding of the leucine codon CUG in several *Candida* species. *Nucleic Acids Research* **21**(17), 4039–4045.

Rosandić, M., Paar, V., 2014. Codon sextets with leading role of serine create "ideal" symmetry classification scheme of the genetic code. *Gene* **543**(1), 45–52.

Rosandić, M., Paar, V., Glunčić, M., 2013. Fundamental role of start/stop regulators in whole DNA and new trinucleotide classification. *Gene* 531(2), 184–190.

Rumer, Y.B., 1966. Systematization of codons in the genetic code. *Doklady Akademii Nauk SSSR* **167**(6), 1393–1394. (in Russian).

Schneider, S.U., Leible, M.B., Yang, X.-P., 1989. Strong homology between the small subunit of ribulose-1,5-bisphosphate carboxylase/oxygenase of two species of *Acetabularia* and the occurrence of unusual codon usage. *Molecular & General Genetics* **218**(3), 445–452.

Seligmann, H., 2000. Evolution and ecology of developmental processes and of the resulting morphology: Directional asymmetry in hindlimbs of Agamidae and Lacertidae (Reptilia: Lacertilia). *Biological Journal of the Linnean Society* **69**(4), 461–481.

Seligmann, H., 2007. Cost minimization of ribosomal frameshifts. *Journal of Theoretical Biology* **249**(1), 162–167.

Seligmann, H., 2010a. The ambush hypothesis at the whole-organism level: Off frame, 'hidden' stops in vertebrate mitochondrial genes increase developmental stability. *Computational Biology and Chemistry* **34**(2), 80–85.

Seligmann, H., 2010b. Do anticodons of misacylated tRNAs preferentially mismatch codons coding for the misloaded amino acid? *BMC Molecular Biology* **11**(41), 1–5.

Seligmann, H., 2011. Error compensation of tRNA misacylation by codon-anticodon mismatch prevents translational amino acid misinsertion. *Computational Biology and Chemistry* **35**(2), 81–95.

Seligmann, H., 2012. Coding constraints modulate chemically spontaneous mutational replication gradients in mitochondrial genomes. *Current Genomics* **13**(1), 37–54.

Seligmann, H., 2015. Phylogeny of genetic codes and punctuation codes within genetic codes. *Biosystems* **129**, 36–43.

Seligmann, H., Pollock, D.D., 2004. The ambush hypothesis: Hidden stop Ccodons prevent off-frame gene reading. *DNA and Cell Biology* **23**(10), 701–705.

Sella, G., Ardell, D.H., 2006. The coevolution of genes and genetic codes: Crick's frozen accident revisited. *Journal of Molecular Evolution* **63**(3), 297–313.

Shcherbak, V.I., 1988. The co-operative symmetry of the genetic-code. *Journal of Theoretical Biology* **132**(1), 121–124.

Shcherbak, V.I., 1989. Rumer's rule and transformation in the context of the co-operative symmetry of the genetic-code. *Journal of Theoretical Biology* **139**(2), 271–276.

Shu, J.-J., Wang, Q.-W., Yong, K.-Y., 2011. DNA-based computing of strategic assignment problems. *Physical Review Letters* **106**(18), 188702.

Shu, J.-J., Wang, Q.-W., Yong, K.-Y., 2016. Programmable DNA-mediated decision maker. *International Journal of Bio-Inspired Computation* **8**(6), 1–5.

Shu, J.-J., Wang, Q.-W., Yong, K.-Y., Shao, F., Lee, K.J., 2015. Programmable DNA-mediated multitasking processor. *Journal of Physical Chemistry B* **119**(17), 5639–5644.

Watson, J.D., Crick, F.H.C., 1953. Molecular structure of nucleic acids - A structure for deoxyribose nucleic acid. *Nature* **171**(4356), 737–738.

Wong, J.R., Lee, K.J., Shu, J.-J., Shao, F., 2015. Magnetic fields facilitate DNA-mediated charge transport. *Biochemistry* **54**(21), 3392–3399.

Yang, C.M., 2004. On the 28-gon symmetry inherent in the genetic code intertwined with aminoacyl-tRNA synthetases—The Lucas series. *Bulletin of Mathematical Biology* **66**(5), 1241–1257.

Yokobori, S.-I., Ueda, T., Watanabe, K., 1993. Codons AGA and AGG are read as glycine in ascidian mitochondria. *Journal of Molecular Evolution* **36**(1), 1–8.

Table of Contents (TOC) Graphic

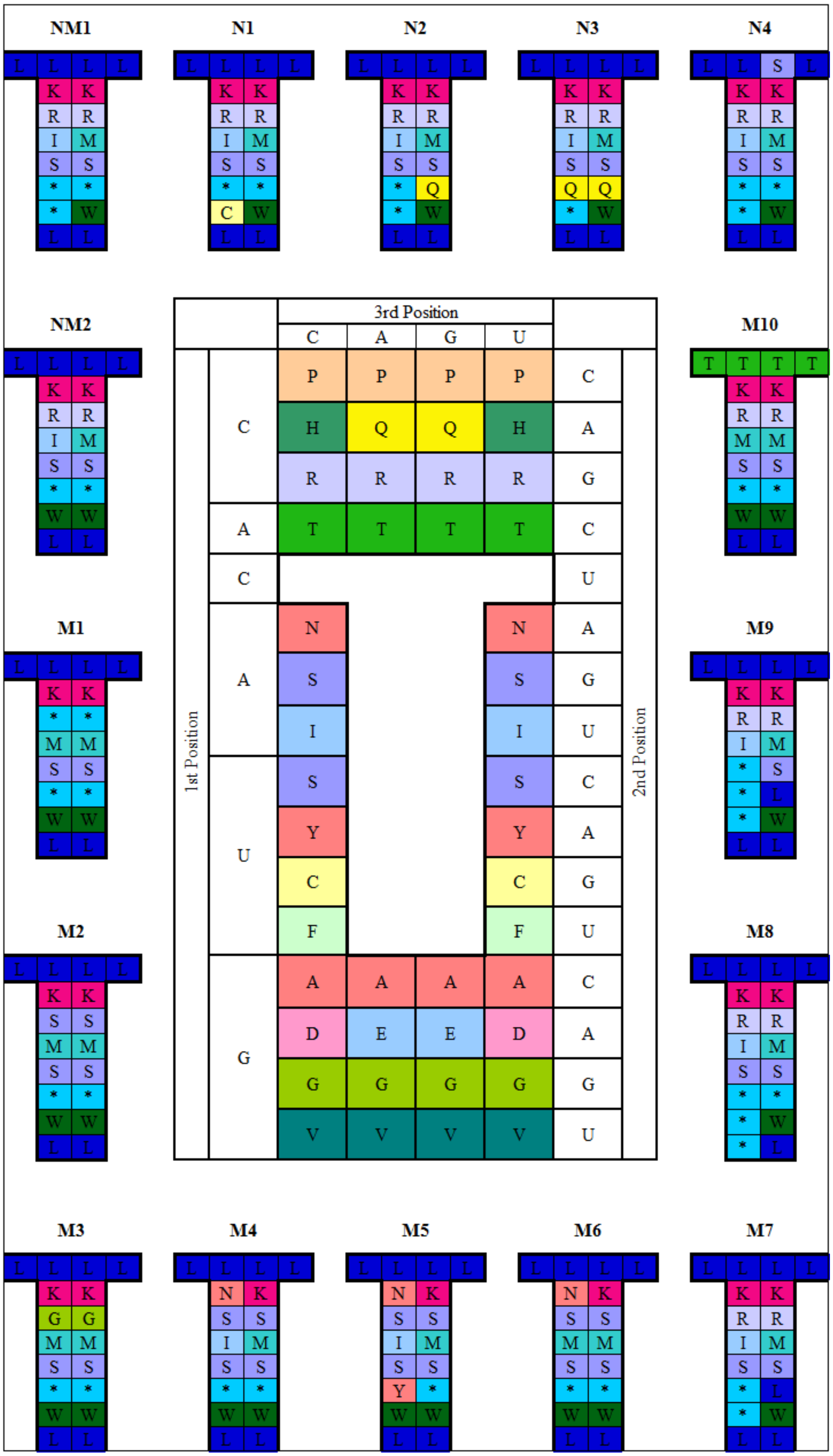